\documentstyle{article}

 \newfont{\goth}{cmbxti10 scaled\magstep1}
 \newfont{\gothi}{cmbxti10}

 \newcommand{\gotig}{\mbox{\gothi g}}

 \newcommand{\gotigh}{\mbox{\gothi h}}

\begin{document}
\title{The Extension Structure of 2D Massive Current Algebras
       }
\author{J. Laartz \thanks{Permanent address :
                   Universit\"at Freiburg,
                   Fakult\"at f\"ur Physik,\newline
                   Hermann-Herder Strasse 3,
                   D-7800 Freiburg i.Br. / Germany}}

 \date{Department of Mathematics, Harvard University\\
      One Oxford Street, Cambridge, MA 02138 / USA \vspace{2em}}

\maketitle
\thispagestyle{empty}
\begin{abstract}
\noindent
The extension structure of the 2-dimensional current algebra of
non-linear sigma models is analysed by introducing Kostant Sternberg $(L,M)$
systems. It is found that the algebra obeys a two step extension by abelian
ideals. The second step is a non-split extension of a representation of the
quotient of the algebra by the first step of the extension. The
cocycle which appears is analysed.
\end{abstract}

\vfill

 \begin{flushright}
 \parbox{12em}
 { \begin{center}
 Harvard University \\
 HUTMP--92/B323 \\
 University of Freiburg \\
 THEP 91/21 \\
 December 1991
 \end{center} }
 \end{flushright}

\newpage
\setcounter{page}{1}

\section{Introduction}
Two-dimensional sigma models whose target space is a symmetric space are
known to be integrable \cite{EF}. They are classically conformal
invariant but after quantization the conformal invariance is broken and
the theory becomes massive. \newline
Recently we have discovered a new current algebra structure for
two-dimensional non-linear sigma models whose target space is a homogenous
space \cite{FLS}. It was found that the algebra closes after introducing a
scalar field with values in the second symmetric tensor power of the
corresponding Lie algebra. In \cite{BFLS} we used this result to analyse
the integrable structure of sigma models on symmetric spaces which belong to
the class of non-ultra-local models.
We derived a
new Lie-Poisson structure differing from the Yang-Baxter structure
obtained in the ultra-local case.\newline
There is a growing intrest in massive current algebras, because massive
theories whose ultraviolet fixed points are conformal theories can
be seen as integrable pertubations of conformal field theories \cite{BPZ}.
In the analysis of the quantum structure of massive current algebras
the conserved non-local charges have been used in \cite{B} to give an
algebraic (non pertubative) definition of a subclass of massive current
algebras.\newline
Another way to achieve a better understanding of the quantized structure of
integrable massive theories would be to construct unitary representations of
the associated classical current algebra. But  for all of these
'Ans\"atze' a better understanding of the algebraic structure of this type
of algebras is needed.

In this note we will deepen the analysis of the algebraic structure of the
massive current algebra derived in \cite{FLS} ( with special emphasis on the
extension structure).
In the second part we will briefly review the results obtained in \cite{FLS}.
In the third part we will go back to an general interpretation of current
algebras in terms of certain modules called $(L,M)$--modules derived by
Kostant and Sternberg in the late 60's while looking at Lie-superalgebras
(partially published in \cite{N}).
We will give a short explanation of this construction and then find an
interpretation of the current algebra derived in \cite{FLS} in terms of a
$(L,M)$--module.
Using this interpretation we will analyse the extension structure of the
current
algebra in part four. We will see that the appearing Schwinger term has an
interpretation (on the level of representations of the quotient of the
whole module by the first extension step) as a cocycle of the second step of an
abelian
extension of this module.

\section{The current algebra of non-linear sigma models on homogenous
         spaces}
We begin by briefly reviewing the results on the current algebra
structure of general non-linear sigma models defined on Riemannian
homogenous spaces $M = G/H$, with action
\begin{equation}
S~=~{\textstyle{1\over 2}} \int d^2\! x \; g_{ij}(\varphi) \,
    \partial^\mu \varphi^i \, \partial_\mu\varphi^j~~~,            \label{eq:S}
\end{equation}
derived in \cite{FLS}. Here $\phi$ is the basic field with values in $M$ and
$g_{ij}$ the metric on $M$. Technically, we require that $M$ is the
quotient space of some connected Lie group $G$, with Lie algebra ${\gotig}$,
modulo some compact subgroup $H\subset G$, with Lie algebra
${\gotigh}\subset{\gotig}$.Then $G$ acts on $M$ by isometries and the $G$
invariance of the action (\ref{eq:S}) leads to a conserved Noether current
$j_\mu$, with values in $\gotig^\ast$ (the dual of $\gotig\,$). The Poisson
algebra of these currents can be written in a closed form after
introducing a scalar field $j$ with values in the second symmetric tensor
power $S^2(\gotig^*)$ of $\gotig^\ast$. To do so it is convenient to introduce
a basis
$(T_a)$ of ${\gotig}$, with structure constants $f_{ab}^c$ defined by
$[T_a,T_b] = f_{ab}^c T_c$, together with the corresponding dual basis
$(T^a)$ of $\gotig^\ast$, and to expand $j_\mu$ and $j$ into components:
$$
j_\mu = j_{\mu,a}T^a \;\;,\;\;\; j = j_{ab}T^a \vee T^b.
$$
With this notation the, the massive current algebra takes the form
\cite{FLS}
\begin{eqnarray}
 \{ j_{0,a}(x) , j_{0,b}(y) \}
 &=& - \, f_{ab}^c \, j_{0,c}(x) \, \delta(x-y)~~,      \label{eq:CA1} \\[1mm]
 \{ j_{0,a}(x) , j_{1,b}(y) \}
 &=& - \, f_{ab}^c \, j_{1,c}(x) \, \delta(x-y) \,
     + \, j_{ab}(y) \, \delta^\prime(x-y)~~,            \label{eq:CA2} \\[1mm]
 \{ j_{0,a}(x) , j_{bc}(y) \}
 &=& - \left( f_{ab}^d \, j_{cd}(x) + f_{ac}^d \, j_{bd}(x) \right)
                                    \delta(x-y)~~,      \label{eq:CA3} \\[2mm]
 \{ j_{1,a}(x) , j_{1,b}(y) \} &=& \, 0~~,              \label{eq:CA4} \\[1mm]
 \{ j_{ab}(x) , j_{cd}(y) \} &=& \, 0~~.                \label{eq:CA6} \\[1mm]
 \{ j_{1,a}(x) , j_{bc}(y) \} &=& \, 0~~,               \label{eq:CA5}
\end{eqnarray}

\section{The massive current algebra as an example for
  a Kostant Sternberg $(L,M)$--system}

\subsection{Kostant Sternberg $(L,M)$--systems}
We begin with a short summary of the notion of $(L,M)$--systems
\cite{KS},\cite{N}. We like to thank S. Sternberg for pointing out this
interesting structure to us and for his unpublished notes \cite{KS} on
which this brief summary is based.\newline
Let $L$ be a Lie algebra and let $M$ be a commutative ring. Let $L$ act on $M$
such
that $L$ is a module over $M$ such that the module structure is consistent with
the representation of $L$ on $M$ and the adjoint representation. That is,
the two linear maps :
\begin{eqnarray}
L \otimes M &\rightarrow& M , \;\;\;\; X \otimes f \mapsto Xf
                                                         \label{eq:lm1}\\
L \otimes M &\rightarrow& L , \;\;\;\; X \otimes f \mapsto fX
                                                         \label{eq:lm2}
\end{eqnarray}
(with $X \in L,\; f \in M$), have to satisfy the following conditions:\newline
%\begin{itemize}
$L$ acts as derivation on $M$,
\begin{equation}
X(fg) = (Xf)g + f(Xg)\;\;;                                 \label{eq:co1}
\end{equation}
the  action of $L$ on $M$ gives a representation of $L$ on $M$,
\begin{equation}
[X,Y](f) = X(Yf) - Y(Xf)\;\;;                               \label{eq:co2}
\end{equation}
$L$ has to be a module over $M$,
\begin{equation}
(fg)X = f(gX)\;\;;                                           \label{eq:co3}
\end{equation}
the module structure is consistent with the action (representation),
\begin{equation}
f(Xg) = (fX)g\;\;;                                          \label{eq:co4}
\end{equation}
the module structure is consistent with the adjoint representation,
\begin{equation}
[X,fY] = (Xf)Y + f[X,Y]\;\;.                                 \label{eq:co5}
\end{equation}

For examples let us start with a closer look at the easiest case the
exterior algebra $A = \bigwedge_M L$ of $L$
as module over $M$ and considering only the zeroth and first order in
the exterior product. $A^0 = M$ and $A^1 = L$ clearly fullfill the above
condition. Furthermore  we have an
additional algebraic structure, that of an anti-Poisson algebra.\newline
Generally an anti-Poisson algebra is a Lie super-algebra $A =
A_0 \oplus A_1$ obeying :
\begin{enumerate}
\item $\{A_1,A_1\} \subset A_1$
\item $\{A_1,A_0\} \subset A_0$
\item $\{A_0,A_0\} \subset A_1$
\end{enumerate}
(opposite parity as in the Poisson algebra case)

\noindent That is: we have a $Z$ and $Z_2$ graded algebra with compatible
$Z$ and
$Z_2$ grading:\newline
$~~ A_0 = A^0 \oplus A^2  \oplus \cdots$\newline
$~~ A_1 = A^1 \oplus A^3  \oplus  \cdots$\newline
in our example $A^0$ is, by construction, commutative, so we
remain with:
\begin{eqnarray}   \label{eq:ap1}
\{ A^1,A^1\} &\subset& A^1\\
\{A^1,A^0\} &\subset& A^0 \label{eq:ap2}
\end{eqnarray}
The definition of $\{.,.\}$ can be extended
to all of $A$ (if $A$ is freely generated by $A^0$ and $A^1$), so that $A$
itself is an anti-Poisson-algebra.\newline
Another example is the space of smooth sections of the exterior bundle
of the tangent bundle of a differentiable manifold $N \;,\;\;\Gamma (\bigwedge
T(N))$. If we take $A^0$ to be the ring of smooth functions on $N$ and
$A^1$ the
Lie algebra of smooth vector fields on $N$, we have again an $(L,M)$--system
and the exterior algebra is an anti-Poisson algebra.\newline
Furthermore the tensor product of two anti-Poisson algebras is again an
anti-Poisson algebra and in the case of a commutative $A_0$, by identifying
the zeroth order term in the tensor product with $M$ and the first order
term with $L$ the tensor product of two $(L,M)$-systems is again an
$(L,M)$-system. We will return to this construction later in more detail.

\noindent The crucial observation of Kostant and Sternberg while looking at
local current commutation relations in the late 60's was that these
relations form an anti-Poisson algebra, which itself could be written in
terms of a tensor product of two
anti-Poisson algebras where one factor is $\Gamma (\bigwedge T(N)
)$ and the second a finite dimensional $(L,M)$-system. \newline
The great advantage by interpreting local current algebras in terms of
$(L,M)$-systems lies in the fact that one gets a depuzzeling into an (easy
to handle, while always the same) infinite dimensional part and a finite
dimensional part characterizing the current algebra. Furthermore the
$M_1$ part also takes care of the distribution as it plays the role of test
functions.\newline
To be more concrete let us write this tensor product explicitly:
\begin{eqnarray}
A_0 = M = M_1 \otimes M_2 &=& C^\infty(N) \otimes M_2~~,\\
A_1 = L = M_1 \otimes L_2 \oplus L_1 \otimes M_2 &=&
            C^\infty(N) \otimes L_2 \oplus {\cal X}(N) \otimes M_2~~,
\end{eqnarray}
As the resulting $(L,M)$--system is again an anti-Poisson algebra it is
useful to write down a multiplication table for the terms of degree zero
and one by using the defining relations (\ref{eq:co1} -- \ref{eq:co5}) and
(\ref{eq:ap1},\ref{eq:ap2}):
\begin{eqnarray}
(f \otimes \alpha).(g \otimes \beta) &=& fg \otimes \alpha\beta
                                                              \label{eq:m1}\\
(f \otimes \alpha).(X \otimes \beta + g \otimes \xi) &=& fX \otimes
       \alpha\beta + fg \otimes \alpha\xi                     \label{eq:m2}\\
(X \otimes \alpha + f \otimes \xi).(g \otimes \beta) &=& Xf \otimes
       \alpha\beta + fg \otimes \xi\alpha                     \label{eq:m3}\\
{}~[X \otimes \alpha , Y \otimes \beta]&=& [X,Y]\otimes \alpha\beta
                                                              \label{eq:m4}\\
{}~[f \otimes \xi , g \otimes \zeta] &=& fg \otimes [\xi , \xi]
                                                              \label{eq:m5}\\
{}~[X \otimes \alpha  , f \otimes \xi] &=& Xf \otimes \alpha\xi - fX
        \otimes \xi\alpha                                     \label{eq:m6}
\end{eqnarray}
where $f,g \in M_1,\;\; X,Y \in L_1,\;\; \alpha , \beta \in M_2$ and $\zeta
,\xi \in L_2$.

\subsection{The current algebra of the non linear sigma model}

For the current algebra of the two dimensional non-linear sigma model on
Riemannian symmetric spaces \cite{FLS} we identify the infinite dimensional
$(L_1,M_1)$
system with
$L_1 = {{\cal X}} (S^1)$ and $M_1 = C^\infty (S^1)$. The finite
dimensional system $(L_2,M_2)$ is given by $L_2 = {\gotig}  \times S^2({\gotig
})$.
(In this semidirect product ${\gotig} $ acts on $S^2({\gotig})$ in the
adjoint  representation) and $M_2 = {\gotig}$.
The multiplication in $M_2$ is trivial (to get a clear distinction between the
two different roles played by $\gotig$ we denote $M_2 = {\gotig}$ by $ \tilde
{\gotig}$) The two linear maps
(\ref{eq:lm1},\ref{eq:lm2}) are given by:
\vfill\eject

\begin{eqnarray}
L_2 \otimes M_2 &\rightarrow& M_2~~,
            \label{eq:si1}\\
(\xi , \alpha) &\mapsto& [\xi,\alpha]~~,\nonumber\\
(\tau , \alpha) &\mapsto& 0  ~~,\nonumber
\end{eqnarray}
The representation of $L_2$ on $M_2$ is the adjoint representation.
\begin{eqnarray}
L_2 \otimes M_2 &\rightarrow& L_2  ~~,                       \label{eq:si2}\\
(\xi , \alpha) &\mapsto& \alpha \vee \xi ~~,\nonumber\\
(\tau , \alpha) &\mapsto& 0 ~~,\nonumber
\end{eqnarray}
so $M_2$ just shifts in $L_2$.
($\xi \in {\gotig}\subset L_2 ,\; \tau \in S^2({\gotig}) \subset
L_2,\; \alpha \in {\gotig} = M_2$).\newline
We can now write the three generators of the current algebra (\ref{eq:CA1} --
\ref{eq:CA6}) as symbols for
certain classes of elements in the $(L,M)$--system.
\begin{enumerate}
\item $<j_0 : f \otimes \xi>$
\item $<j_1 : X \otimes \alpha>$
\item $<j : f \otimes \tau>$
\end{enumerate}
For the current algebra in this version we derive from
(\ref{eq:m1} -- \ref{eq:m6}) by using (\ref{eq:si1}) and (\ref{eq:si2}):
\begin{eqnarray}
{}~[ <j_0 : f \otimes \xi+> , <j_0 : g \otimes \zeta> ] &=&
              <j_0 : fg \otimes [\xi , \zeta]> \label{eq:lmp1}\\
{}~[ <j_0 : f \otimes \xi> , <j_1 : X \otimes \alpha> ] &=&
              <j_1 : fX \otimes [\xi , \alpha]>  - \nonumber\\
           ~&&~~~~~~~~~~~~- <j : Xf \otimes (\alpha \vee \xi)>\label{eq:lmp2}\\
{}~[ <j_0 : f \otimes \xi> , <j : g \otimes \tau> ] &=&
              <j : fg \otimes [\xi , \tau]> ~~.\label{eq:lmp3}
\end{eqnarray}
\section{The Extension Structure}

To further analyse this algebraic structure let us write the tensor product
of the two $(L,M)$--systems explicitly:
\begin{eqnarray}
M = M_1 \otimes M_2 &=& C^\infty (\Sigma) \otimes \tilde{\gotig}\\
L = L_1 \otimes M_2 \oplus M_1 \otimes L_2 &=& {\cal X}(S^1) \otimes
\tilde{\gotig} \otimes C^\infty (S^1) \otimes {\gotig} \times S^2{\gotig}
\label{eq:te2}
\end{eqnarray}
Let us begin with a few observations:
\begin{itemize}
\item Because of (\ref{eq:lmp2}), we notice that there is a `twist' in the
sum of (\ref{eq:te2}), that is, the sum is a semidirect sum.
\item Because of (\ref{eq:lmp1}), $M_1 \otimes {\gotig}$ is a submodule.
\item Furthermore we see that we have an extension of $M_1 \otimes {\gotig}$ by
the abelian ideal $L_1 \otimes M_2 \oplus M_1 \otimes S^2(\gotig)$.
\end{itemize}
Actually this is not the whole story, as a result of (\ref{eq:lmp2}) there is a
two step extension by abelian ideals. Namely:
\begin{equation}
\begin{array}{c}
M_1\otimes S^2{\gotig}\\
\cap\\
L_1 \otimes M_2 \oplus M_1 \otimes S^2{\gotig}\\
\cap\\
L_1 \otimes M_2 \oplus M_1 \otimes L_2~.
\end{array} \label{eq:ex}
\end{equation}
So we have the following structure:
\begin{enumerate}
\item The quotient of $L$ by the first extension is $M_1 \otimes {\gotig}$.
\item The quotient of the first extension by the second is $L_1 \otimes M_2$.
      So $L_1 \otimes M_2$ itself is not an ideal but rather it is the cokernel
      of the second extension.
\item Relations (\ref{eq:lmp1} -- \ref{eq:lmp3}) can be read as a
      representation of $M_1 \otimes {\gotig}$ on $L$, so that in particular
      we have a representation on the extensions.
\end{enumerate}
To make the above observations more concrete, we look at the short exact
sequence
determining the first step of the extension:
$$
0 \rightarrow V^\prime \rightarrow V \rightarrow V^{\prime\prime}
\rightarrow 0~~,
$$
where
\begin{itemize}
\item $V^\prime = L_1 \otimes M_2 \oplus M_1 \otimes S^2{\gotig} ~~,$
\item $V = L_1 \otimes M_2 \oplus M_1 \otimes L_2~~,$
\item $V^{\prime\prime} = M_1 \otimes {\gotig}~~.$
\end{itemize}
The first step of the extension is charcterized by the corresponding long
exact sequence of cohomologies of $M_1 \otimes {\gotig} = \Xi$ (as a Lie
algebra) with values in the corresponding
submodules:
$$
\rightarrow H^1({\Xi}, V^{\prime\prime}) \rightarrow H^2({\Xi},V^\prime)
\rightarrow H^2({\Xi},V) \rightarrow H^2({\Xi},V^{\prime\prime})
\rightarrow~~~.
$$
Because of (\ref{eq:lmp1}) the cohomology class of $H^2({\Xi},V) $ is zero,
so that we have a split extension, i.e. no cocycle appears.\newline
The second step of the extension is less trivial. To analyse it, recall
the observation that the relations (\ref{eq:lmp1} -- \ref{eq:lmp3}) can be
read as representaions of $M_1 \otimes {\gotig}$ on the subsequent
representation spaces. So we will look
at the short exact sequence of representations of $M_1 \otimes {\gotig}$ :
$$
0 \rightarrow W^\prime \rightarrow W \rightarrow W^{\prime\prime}
\rightarrow 0
$$
where
\begin{itemize}
\item $W^\prime = M_1 \otimes S^2{\gotig} $
\item $W = L_1 \otimes M_2 \oplus M_1 \otimes S^2{\gotig}$
\item $W^{\prime\prime} = L_1 \otimes M_2$
\end{itemize}
We want to calculate the extensions
of $W^{\prime\prime}$ by $W^\prime$ as representations of $\Xi$, which are
described by the functor $Ext$. So we should analyse
$Ext^1_{\Xi}(W^{\prime\prime},W^\prime)$.
If we look again at (\ref{eq:lmp2}) we see that this tells us that we have a
splitting (now a splitting of vector spaces) of the short exact sequence by $X
\otimes \alpha$ and the cocycle should be given by ``$Xf\otimes
\alpha \vee\xi$''. Let us calculate this cocycle explicitly. \newline
To derive the cocycle condition we look at the given splitting of the short
exact sequence:
$$
\begin{array}{ccccccccc}
0& \rightarrow &W^\prime &\rightarrow& W &\rightarrow &W^{\prime\prime}&
\rightarrow& 0\\
{}~&~&~&~&~&\hookleftarrow&~&~\\
{}~&~&~&~&~&\phi (split)&~&~
\end{array}
$$~
To be a splitting $\phi$ has to be an algebra homomorphism,
$$
g \phi (W^{\prime\prime}) = \phi(g W^{\prime\prime}) + C_g(W^{\prime\prime})~~.
$$
This means that from
$$
[g,g^\prime] \nu = g(g^\prime \nu) - g^\prime (g\nu)
$$
we derive the cocycle condition by evaluating this for $\nu = \phi(W'')$
and $g,g'\in \Xi$,~
\begin{equation}
[g,g^\prime] (W^\prime + \phi(W^{\prime\prime})) - g(g^\prime( W^\prime +
\phi(W^{\prime\prime})) + g^\prime (g W^\prime + \phi(W^{\prime\prime})) =
0~~.\nonumber
\end{equation}
The cocycle condition reads:
\begin{equation}
C_{[g,g^\prime]} (W^{\prime\prime}) + g C_{g^\prime} (W^{\prime\prime}) -
C_{g^\prime}(g W^{\prime\prime}) + C_g(g^\prime W^{\prime\prime}) -
g^\prime C_g (W^{\prime\prime}) = 0 ~~.\label{eq:cocy1}
\end{equation}
This means for $g = f \otimes \xi$, $s = X \otimes \alpha  \in W''$,
and $C_g(s) = Xf \otimes \xi \vee \alpha$:
$$
X(fg) \otimes [\xi,\eta]\vee\alpha - (fXg + gXf)\otimes
[\xi,\eta]\vee\alpha = 0~~.
$$
Which is exactly the requirement that $L_1$ acts by derivation on $M_1$
(but preserving the tensor product structure).
As this is a cocycle in the tensor product we can ask whether the two factors
are
cocycles seperatly. Therefore we look at the two components of the map
from
\begin{eqnarray*}
\Xi \times W''& \rightarrow& W'\\
(M_1\otimes\gotig)\times(L_1\otimes M_2) &\rightarrow& (M_1\otimes
S^2(\gotig))\\
((f\otimes\xi),(X\otimes\alpha))&\mapsto&
Xf\otimes(\xi\vee\alpha)\\
\end{eqnarray*}
independently.
We first look at $M_1\otimes L_1 \rightarrow M_1 $,
given by $ (f,X) \mapsto Xf$
We would like to know whether this defines a cocycle. So we analyse the
map \newline $M_1 \hookrightarrow M_1 + L_1$ \newline given by
$$<f,g + X> = fg + fX + Xf.$$
Using (\ref{eq:cocy1}) for this homomorphism one obtains:
$$
X(ff') - f'Xf - fXf' = 0~.
$$
This is the Leibniz rule for $L_1$, so $Xf$ is indeed a cocycle.\newline
For the second part $\gotig \otimes M_2 \rightarrow S^2(\gotig)$, given by
$(\xi,\alpha) \mapsto \xi \vee \alpha$, we look at the
map\newline
$\gotig\hookrightarrow M_2 + S^2(\gotig)$ \newline defined by
$$<\xi,\sigma + \alpha> = ad_\xi\sigma + ad_\xi\alpha + \xi \vee\alpha$$
using again (\ref{eq:cocy1}) to obtain the cocycle condition we have
$$<[\xi,\xi'],\sigma + \alpha> - <\xi,<\xi',\sigma + \alpha>> + <\xi',<\xi,
\sigma + \alpha>> = [\xi,\xi'] \neq 0$$
So this is not a cocycle.

\noindent Let us summarize the result on the
cocycle structure: the tensor product of the above
maps defines a 1-cocycle, which comes from a 1-cocycle defined by the
action of the vector fields on the ring of functions over a manifold. This
cocycle is converted into the 1-cocycle of the tensor map with the help of
a homomorphism defined by the second map.

\noindent {\bf Remarks:}
\begin{enumerate}
\item
This structure gives us information about the functor $Ext^1_\Xi$, because
we see now that the interesting contribution to $Ext$ comes from one of the
associated cup products:\newline
$Ext^1_{M_1}(L_1,M_1) \otimes Hom_{\gotig} ({\gotig} \otimes M_2,
S^2({\gotig}))
\rightarrow Ext^1_\Xi(L_1\otimes M_2,M_1\otimes S^2({\gotig}))$.\newline
This cup product describes exactly the tensor product of the cocycles
defined by the first
map with ${\gotig}$-equivariant homomorphisms ${\gotig} \otimes
M_2,\rightarrow S^2({\gotig})$ which are 1-cocycles of the tensor product of
maps.
\item The above construction is also possible for arbitrary
representation spaces of ${\gotig} \subset L_2$. By this we mean that it is
possible to replace $S^2{\gotig}$ and $M_2$ by other representation spaces.
\item One would expect to get more information by using the fact that we
are only interested in the $M$-invariant part of $Ext^1$,
that is, constructing a double complex and looking at the long exact
sequence:
$$
0 \rightarrow H^1(M , Hom_{\Xi}) \rightarrow
Ext^1_{(M,{\Xi})} \rightarrow (Ext^1_{\Xi})^{M_{inv}}
\rightarrow H^2(M, Hom_{\Xi}) \rightarrow~~.
$$
But because of the trivial multiplication in $M_2$, the $M$ invariance of the
construction gives no further restriction on the appearing cocycle.
\end{enumerate}
\vfill\eject

{\bf Acknowledgement:} First of all I would like to thank S.Sternberg for
pointing to me the $(L,M)$--structures and for fruitful
discussions; Furthermore R. Pinck for teaching me homological algebra and
M. Forger, E. Frenckel, V. Kac for fruitful discussions.
The warm hospitallity at Harvard  Mathematics Department is acknowledged.
This work is suported by a grant from the Studienstiftung des Deutschen Volkes.


\begin{thebibliography}{xxxx}
\bibitem{BPZ} A.~Belavin, A.~Polyakov, A.~Zamolodchikov, Infinite
dimensional symmetry in two-dimensional quantum field theory, Nucl.Phys.
B241 (1984) 333

\bibitem{B} D.~Bernard, Hidden Yangians in 2D Massive Current Algebras,
Commun.Math.Phys. 137 (1991) 191

\bibitem{BFLS} M.~Bordemann, M.~Forger, J.~Laartz, U.~Sch\"aper,
The Lie-Poisson Structure of Integrable Classical Non-Linear Sigma Models,
Freiburg preprint THEP 91/11

\bibitem{EF} H.~Eichenherr, M.~Forger, On the Dual Symmetry of the
Non-Linear Sigma Models, Nucl.Phys. B155 (1979) 381

\bibitem{FLS} M.~Forger, J.~Laartz, U.~Sch\"aper, Current Algebra of
Classical Non-Linear Sigma Models, Freiburg preprint THEP 91/10, to appear in
Commun.Math.Phys. 145 (1992)

\bibitem{KS} B.~Kostant, S.~Sternberg, Anti-Poisson Algebras and Current
Algebras, Unpublished notes

\bibitem{N} Y.~Ne'eman, Commutateurs de Courants locaux et termes a
gradient dans une algebre de Lie, Conference donnee a Strasbourg CNRS du 3
juin 1971

\end{thebibliography}
\end{document}